\documentclass[final]{aipproc}

\layoutstyle{8x11single}

\begin{document}

\title{Renormalization of One Meson Exchange Potentials \\ 
and Their Currents}

\classification{03.65.Nk,11.10.Gh,13.75.Cs,21.30.Fe,21.45.+v}
\keywords      {NN interaction, One Boson Exchange, Renormalization, Strong form factors, Large Nc , Chiral
symmetry, Gauge invariance.}

\author{\underline{A. Calle Cord\'on~\footnote{Speaker at 
12th International Conference On Meson-Nucleon Physics And The Structure Of The Nucleon (MENU 2010) 31 May - 4 Jun 2010, Williamsburg, Virginia}}\,\,}{
  address={Departamento de F\'isica At\'omica, Molecular y Nuclear, Universidad de Granada, E-18071 Granada, Spain.}
}

\author{\, E. Ruiz Arriola}{
  address={Departamento de F\'isica At\'omica, Molecular y Nuclear, Universidad de Granada, E-18071 Granada, Spain.}
}

\begin{abstract}

The Nucleon-Nucleon One Meson Exchange Potential, its wave functions
and related Meson Exchange Currents are analyzed for point-like
nucleons.  The leading $N_c$ contributions generate a local and energy
independent potential which presents $1/r^3$ singularities, requiring
renormalization.  We show how invoking suitable boundary conditions,
neutron-proton phase shifts and deuteron properties become largely
insensitive to the nucleon substructure and to the vector
mesons. Actually, reasonable agreement with low energy data for
realistic values of the coupling constants (e.g. SU(3) values) is
found.  The analysis along similar lines for the Meson Exchange
Currents suggests that this renormalization scheme implies tremendous
simplifications while complying with exact gauge invariance at any
stage of the calculation.

\end{abstract}

\maketitle

\section{Introduction}

The original idea of Yukawa that NN interaction at long distances is
due to One Pion Exchange (OPE) was verified quantitatively by the
Nijmegen benchmarking partial wave analysis for NN scattering in the
elastic region with $\chi^2 /{\rm DOF} \sim 1$; a partial wave and
energy dependent square well potential was considered for distances
below 1.4-1.8 fm~\cite{Stoks:1993tb} while the neutral and charged
pion masses could be determined from the fit to their currently
accepted PDG values assuming OPE above such distances. The
verification of other meson exchanges is less straightforward from NN
elastic scattering since the shortest resolution distance probed at
the pion production threshold is about $\lambda = \hbar /\sqrt{M_N
  m_\pi} \sim 0.5 {\rm fm}$ and effectively short distance
interactions admit a variety of parameterizations and
forms~\cite{Stoks:1994wp, Wiringa:1994wb,Machleidt:2000ge} which show
up quantitatively when nucleons are placed off-shell by the presence
of a third particle. Already in the simplest case of a photon as the
additional particle either in the initial or final state (see
\cite{Riska:1989bh} and references therein) Gauge invariance relates
meson exchange potentials and currents (MEC's) but also requires that
Hamiltonian eigenstates are used to compute electroweak matrix
elements. Clearly, one should not expect to know the longitudinal
currents any better than potentials. Ambiguities in transverse,
i.e. non-minimal coupling, terms of the current just reflect the
composite character of Nucleons as well as their finite size. In the
OBE potential, strong form factors are added to mimic this finite
size~\cite{Machleidt:1987hj}, which is about $0.6 {\rm fm}$. In the
case of gauge invariance, the inclusion of a form-factor introduced by
hand, i.e., not computed consistently within meson theory, implies a
kind of non-locality in the interaction which could be made gauge
invariant by introducing link operators between two points, thereby
generating a path dependence, for which no obvious resolution has been
found yet. Note that purely phenomenological potentials not based
entirely on the Meson Exchange picture are inherently ambiguous. If
the corresponding NN wave functions are combined with MEC's
conflicting results with gauge invariance are eventually produced.
Since the finite size of the nucleon is comparable to the minimal
resolution probed in NN scattering, we do not expect to see the
difference between a point-like nucleon and an extended one at
sufficiently low energies.  Recently~\cite{Cordon:2009pj} we have
suggested replacing strong form factors in the NN potential by
renormalization conditions on low energy scattering properties. In the
lowest partial waves we have shown that {\it after} renormalization
finite nucleon effects parameterized as strong form factors are indeed
marginal.  We suggest using renormalization ideas for potentials, wave
functions and currents computed consistently. In the present
contribution we review our findings~\cite{Cordon:2009pj} and apply
them to analyze the radiative neutron capture, $n+p \to d + \gamma$, a
nuclear reaction where the MEC are known to be
essential~\cite{Riska:1989bh}.

\section{Meson Exchange Potentials}

A useful and simplifying assumption arises from our observation that
the symmetry pattern of the sum rules for the old nuclear Wigner and
Serber symmetries discussed in
Refs.~\cite{CalleCordon:2008cz,CalleCordon:2009ps} largely complies to
the large $N_c$ and QCD based contracted $SU(4)_C$
symmetry~\cite{Kaplan:1996rk}. In the large $N_c$ limit with $\alpha_s
N_c$ fixed, nucleons are heavy, $M_N \sim N_c$, and the definition of
the NN potential $\sim N_c$ makes sense.  The tensorial spin-flavour
structure was found to be~\cite{Kaplan:1996rk}
\begin{eqnarray} 
V (r) = V_C (r) + \tau_1 \cdot \tau_2 \left[ \sigma_1 \cdot \sigma_2  W_S (r)
+ S_{12}  W_T(r) \right]  \sim N_c \, . 
\label{eq:pot-largeN}
\end{eqnarray} 
Other operators such as spin-orbit or relativistic corrections are
${\cal O} (N_c^{-1}) $ and hence suppressed by a relative $1/N_c^2$
factor. While these counting rules are directly obtained from
quark-gluon dynamics, quark-hadron duality and confinement requires
that above the confinement scale one can saturate
Eq.~(\ref{eq:pot-largeN}) with multiple exchanges of mesons which have
a finite mass for $N_c \gg 3$~\cite{Banerjee:2001js}. We retain one
boson exchange (OBE) with $\pi$,$\sigma$,$\rho$ and $\omega$ and $a_1$
mesons~\footnote{Other mesons such as $\eta$ are sub-leading.  Due to
the $U_A(1)$ anomaly the $\eta'$ meson appears to be heavy, but in the
large $N_c$ limit one it becomes degenerate with the pion $m_{\eta'} =
m_\pi + {\cal O}(1/N_c)$. So, one might think that this meson would be
as important as the pion itself. Actually this is not so since being
an iso-scalar state it generates terms in the potential
$V_S \sigma_1 \cdot \sigma_2 $ and $V_T S_{12} $ which are ${\cal
O}(1/N_c)$ and hence do not contribute to the leading potential in
Eq.~(\ref{eq:pot-largeN}).}.  The corresponding potential
reads~\footnote{In our previous work we left out the $a_1$ meson. We
use the chiral Lagrangian~\cite{Stoks:1996yj} and take $g_{a_1 NN} =
(m_{a_1}/m_\pi) f_{\pi NN} = 8.4$.}
\begin{eqnarray}
V_C (r) &=& - \frac{g_{\sigma NN}^2}{4 \pi} \frac{e^{-m_\sigma r}}{r}
+ \frac{g_{\omega NN}^2}{4 \pi} \frac{e^{-m_\omega r}}{r}  \, , 
\label{eq:vc} \\ 
W_S(r)
&=& \frac{g_{\pi NN}^2}{48\pi} \frac{m_\pi^2}{\Lambda_N^2}
\frac{e^{-m_\pi r}}{r} + \frac{f_{\rho NN}^2}{24
\pi}\frac{m_\rho^2}{\Lambda_N^2} \frac{e^{-m_\rho r}}{r}  
-\frac{g_{a_1 NN}^2}{ 6 \pi} \frac{e^{-m_{a_1} r}}{r}  
\, , 
\\ 
W_T(r) &=&
\frac{g_{\pi NN}^2}{48 \pi}\frac{m_\pi^2}{\Lambda_N^2} \frac{e^{-m_\pi r}}{r} \left[ 1 + \frac{3}{m_\pi r} + \frac{3 }{(m_\pi r)^2}\right] 
- \frac{f_{\rho NN}^2}{48 \pi}\frac{m_\rho^2}{\Lambda_N^2} \frac{e^{-m_\rho
r}}{r} \left[ 1 + \frac{3}{m_\rho r} + \frac{3 }{(m_\rho r)^2}\right] \nonumber 
\\ 
&+& \frac{g_{a_1 NN}^2}{12 \pi} \frac{e^{-m_{a_1} r}}{r} \left[ 1 + \frac{3}{m_{a_1} r} + \frac{3 }{(m_{a_1} r)^2}\right] 
\, ,  
\end{eqnarray} 
where $\Lambda_N = 3 M_p /N_c$ and $g_{\sigma NN}, g_{\pi NN}, f_{\rho
  NN}, g_{\omega NN}, g_{a_1 NN} \sim \sqrt{N_c}$ and $m_\pi ,
m_\sigma, m_\rho, m_\omega, m_{a_1} \sim N_c^0$. To leading and
sub-leading order in $N_c$ one may neglect spin orbit, meson widths and
relativity.  The tensor force $W_T$ is singular at short distances
$\sim 1/r^3$ and requires renormalization. The renormalization is
carried out in coordinate space using a boundary condition at a short
distance cut-off $r_c$ (see \cite{Cordon:2009pj} for details) which
makes the Hamiltonian self-adjoint for $r > r_c$. Besides being much
simpler and efficient, this method allows to deal with cut-off
independent potentials. In practice convergence is achieved for $r_c
\sim 0.3 {\rm fm}$ (see e.g. left panel Fig.~\ref{fig:mecs} for the
asymptotic D/S ratio$\eta$)~\footnote{Imposing a cut-off in momentum
  space generates an apparent delayed convergence due to the long
  distances distortion of the potential, so that unexpectedly large
  momentum cut-offs are needed.  Renormalized results, however, agree
  in both $r-$ and $p-$spaces~\cite{Valderrama:2007ja}.}.

\begin{table}
\resizebox{1.\textwidth }{!}{%
\begin{tabular}{cccccccccccc} 
\hline
& $\gamma ({\rm fm}^{-1})$ & $\eta$ & $A_S ( {\rm fm}^{-1/2}) $
& $r_m ({\rm fm})$ & $Q_d ( {\rm fm}^2) $ & $P_D $ & $\langle r^{-1}
\rangle $ & $ \alpha_0 ({\rm fm}) $ & $\alpha_{02} ({\rm fm}^3) $ & $
\alpha_2 ({\rm fm}^5) $ & $r_0 ({\rm fm} ) $ \\ \hline
$\pi\sigma\rho\omega a_1$ (AVMD)
 & Input & 0.02557 & 0.8946 & 1.9866 & 0.2792 & 6.53\% & 0.639 & 5.467 & 1.723 &
6.621 & 1.722\\ 
$\pi\sigma\rho\omega a_1$ (PDG)
 & Input & 0.02552 & 0.8937 & 1.9846 & 0.2780 & 6.58\% & 0.671 & 5.463 & 1.714 &
6.607 & 1.712\\ 
$\pi\sigma\rho\omega$$^* a_1$ (AVMD)
 & Input & 0.02544 & 0.8966 & 1.9909 & 0.2788 & 5.90\% & 0.5087 & 5.477 & 1.720
& 6.604 & 1.734\\ 
$\pi\sigma\rho\omega$$^* a_1$ (PDG)
 & Input & 0.02540 & 0.8951 & 1.9876 & 0.2773 & 6.01\% & 0.557 & 5.470 & 1.708 &
6.588 & 1.724\\
\hline 
NijmII & Input & 0.02521 & 0.8845(8) & 1.9675 & 0.2707 & 
5.635\% &  0.4502 & 5.418 & 1.647 & 6.505 & 1.753 \\
Reid93 & Input & 0.02514 & 0.8845(8) & 1.9686 & 0.2703 & 
5.699\% & 0.4515 & 5.422 & 1.645 & 6.453 & 1.755 \\ \hline 
Exp.  &  0.231605 &  0.0256(4)  & 0.8846(9) & 1.9754(9)  &
0.2859(3) & 5.67(4)  & $-$ & 5.419(7) & $-$ & $-$ & 1.753(8) \\ \hline 
\end{tabular}
}
\caption{\label{tab:table_axial} Deuteron properties and low energy
parameters in the $^3S_1-^3D_1$ channel for OBE potentials including
$\pi$,$\sigma$, $\rho$, $\omega$ mesons~\cite{Cordon:2009pj} as well
as axial-vector meson $a_1$. We use the same numbers and notation as
in Ref.~\cite{Cordon:2009pj}. Here AVMD means taking $m_{a_1} =
\sqrt{2} m_\rho \simeq 1107 \rm MeV$ and PDG taking $m_{a_1} =1230
\rm MeV$. See also Ref.~\cite{deSwart:1995ui} and references therein.}
\end{table}

\begin{figure*}[tbc]
\includegraphics[height=5.5cm,width=5.5cm,angle=270]{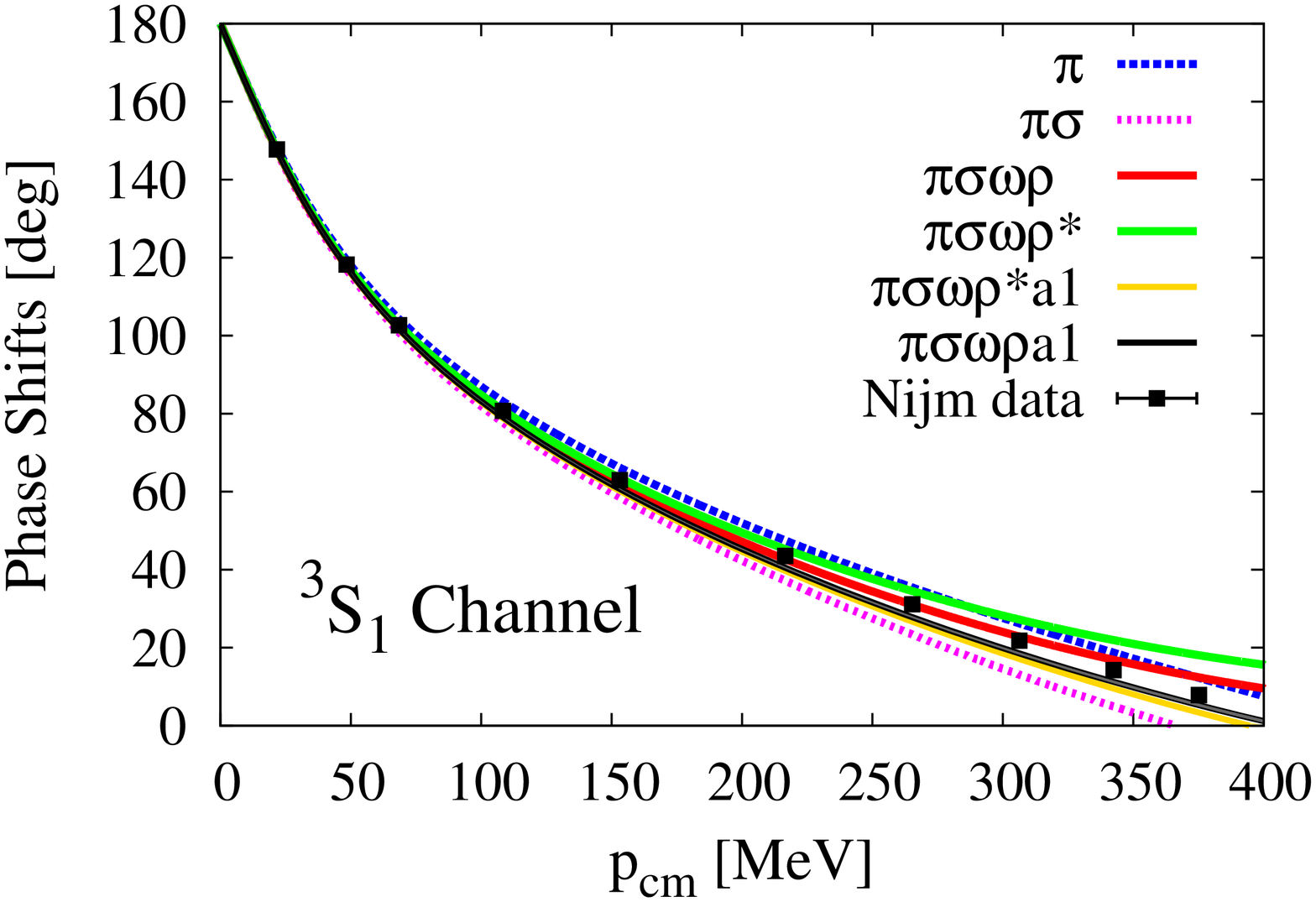}
\includegraphics[height=5.5cm,width=5.5cm,angle=270]{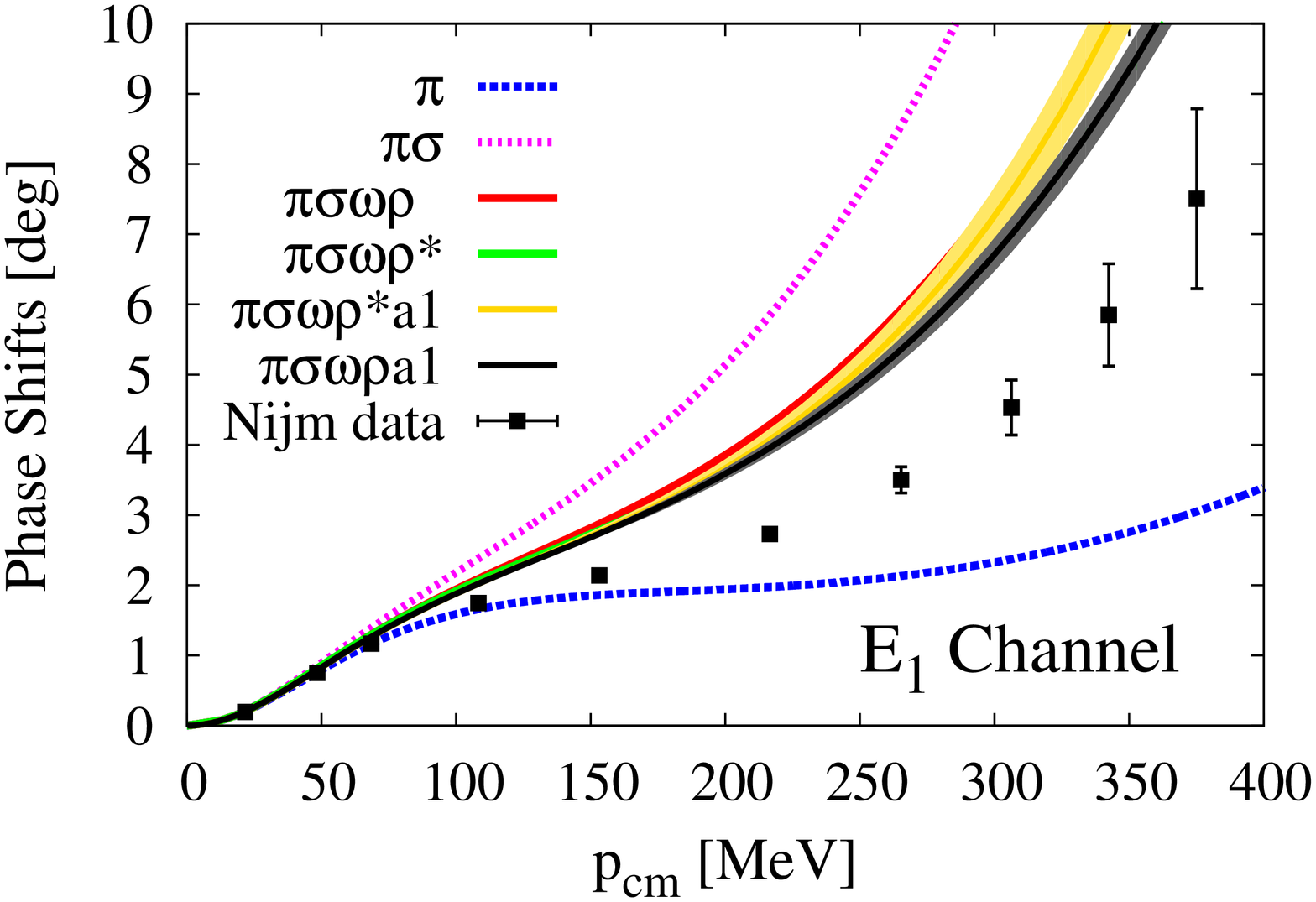}
\includegraphics[height=5.5cm,width=5.5cm,angle=270]{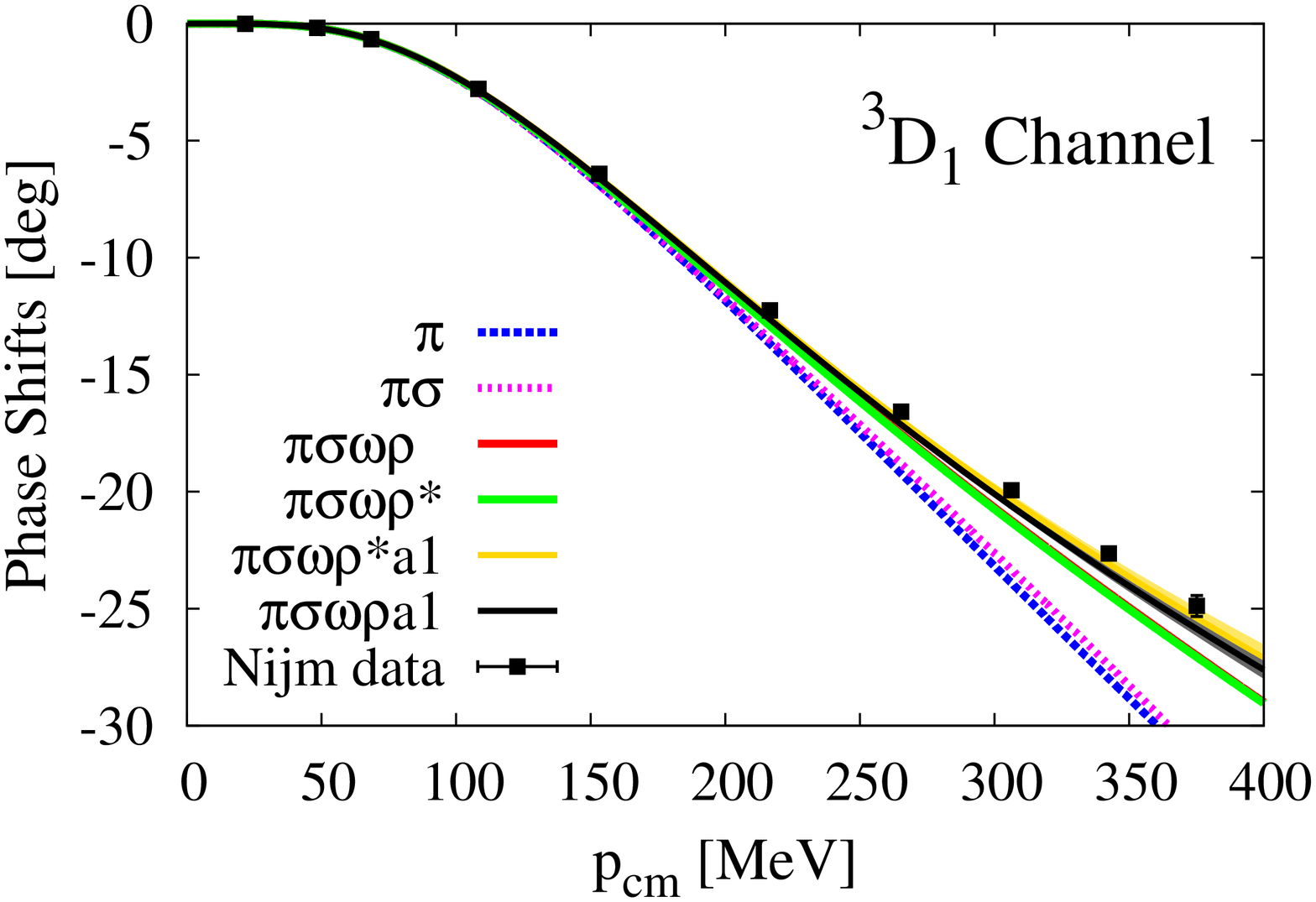}
\caption{$np$ spin triplet eigen phase shifts for the total angular momentum
$j=1$ as a function of the c.m. momentum compared to an average of the
Nijmegen partial wave analysis and high quality potential
models~\cite{Stoks:1994wp}. See table 1 for notation.  The band in the
case of the $a_1$ represents the error of changing $m_{a_1}$ from the
AVMD value to the PDG value.}
\label{fig:phase-triplet-a1}
\end{figure*}

\begin{figure*}[tbc]
\includegraphics[height=5.5cm,width=5.5cm,angle=270]{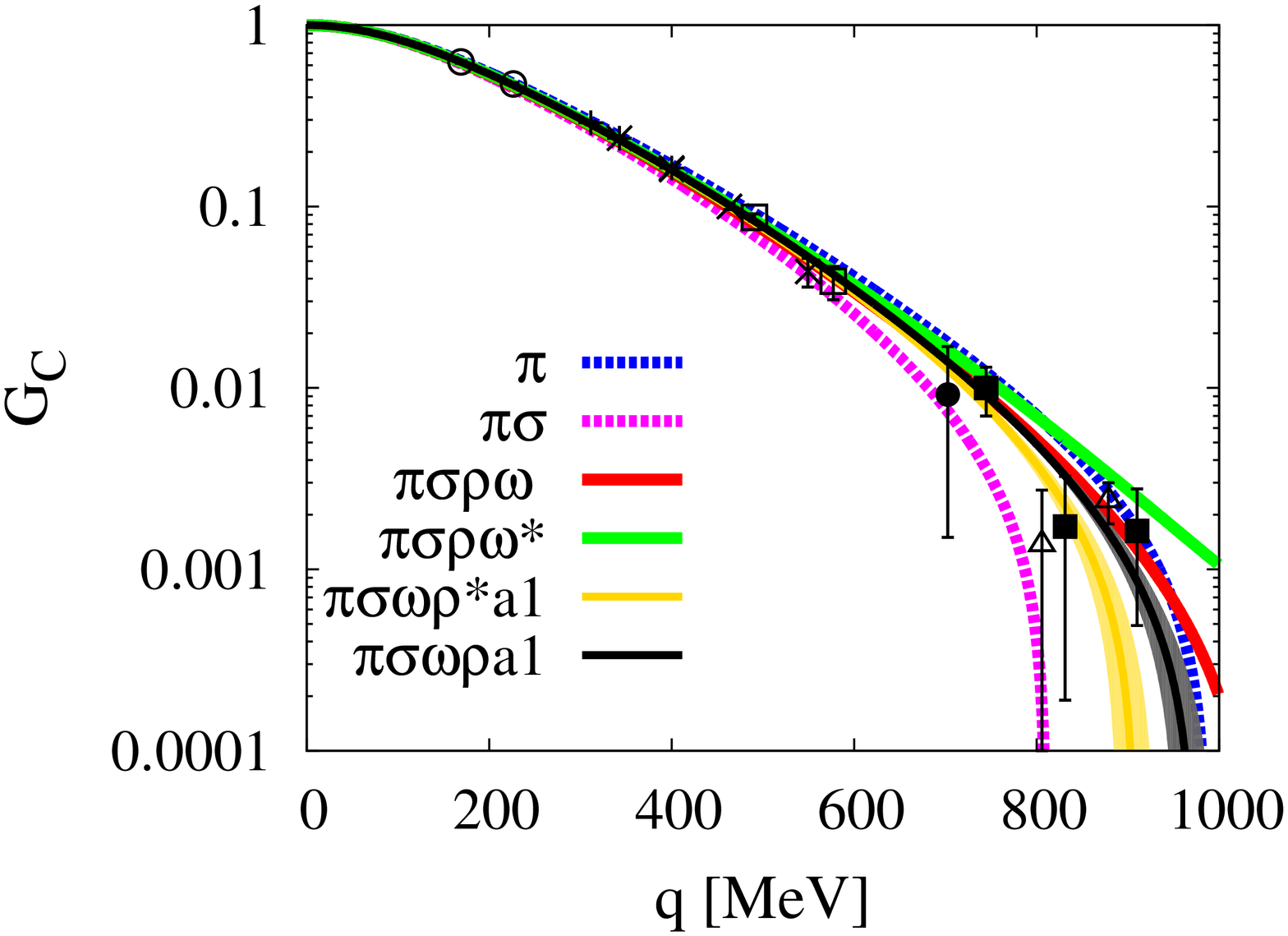}
\includegraphics[height=5.5cm,width=5.5cm,angle=270]{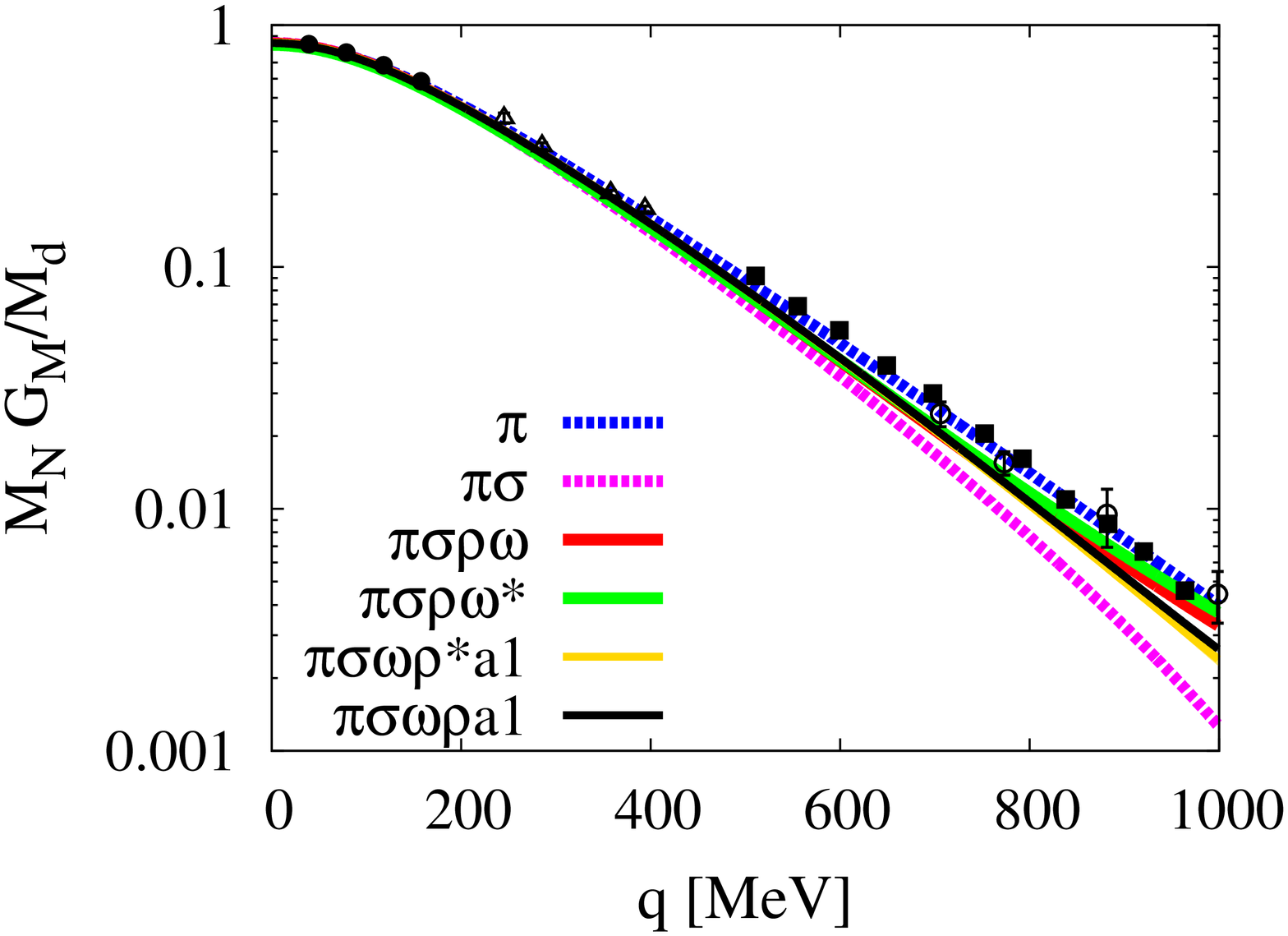}
\includegraphics[height=5.5cm,width=5.5cm,angle=270]{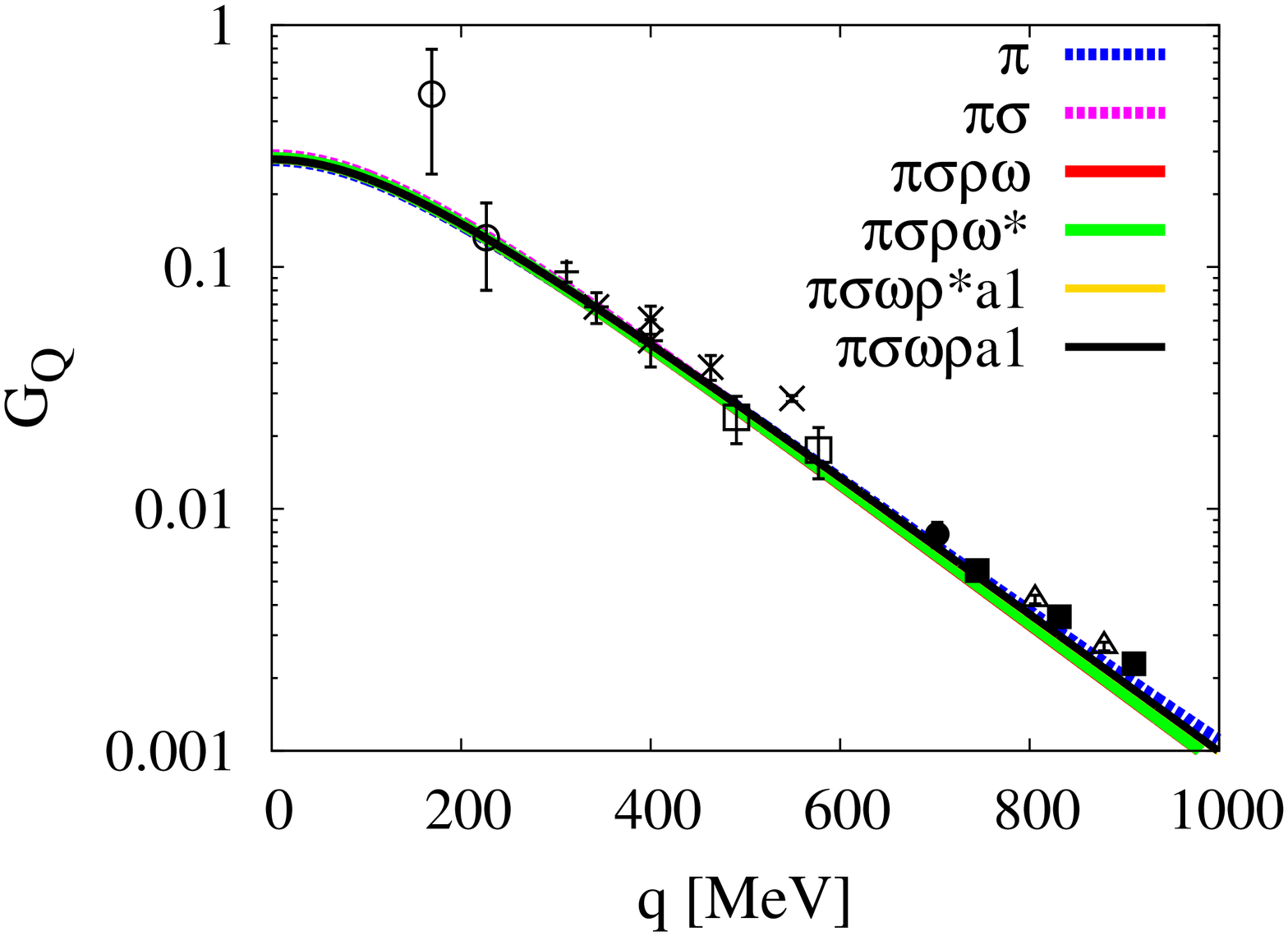}
\caption{Deuteron charge (left), magnetic (middle) and quadrupole
  (right) form factors as a function of transfer $q$ (in MeV).  See
 Table 1 for notation. Data can be traced from
 Ref.~\cite{Gilman:2001yh} (see also references
 in \cite{Valderrama:2007ja}).}
\label{fig:formfactors}
\end{figure*}

Overall, the agreement is good for {\it realistic} couplings and
masses as expected from other sources (see Ref.~\cite{Cordon:2009pj}
for a short compilation) including a natural SU(3) value for
$g_{\omega NN}$ coupling.  The deuteron properties and low energy
parameters are shown in table~\ref{tab:table_axial}.  The $^1S_0$
phase shift is reproduced for $m_\sigma\sim 500 {\rm MeV}$ while the
$^3S_1-^3D_1$ phase shifts are plotted in
Fig.~\ref{fig:phase-triplet-a1}. Space-like electromagnetic form
factors in the impulse approximation~\cite{Gilman:2001yh} for $G_E^p
(-{\bf q}^2)=1/(1+{\bf q}^2/m_\rho^2)^2$ and without MEC are plotted
in Fig.~\ref{fig:formfactors} (see \cite{Valderrama:2007ja} for the
$\pi$ case). As we see, including shorter range mesons induces
moderate changes, due to the expected short distance insensitivity
embodied by renormalization, {\it despite} the short distance
singularity and {\it without} introducing strong meson-nucleon-nucleon
vertex functions~\footnote{We include only the OBE part
  of the leading $N_c$ potential but multiple meson exchanges could
  also be added as well as $\Delta$ degrees of freedom to comply with
  large $N_c$ counting rules~\cite{Banerjee:2001js}.
  Eq.~(\ref{eq:pot-largeN}) yields $V_c$ at leading order in $N_c$. It
  is worth reminding that for chiral potentials $V_c = {\cal O} (g_A^4
  /(f_\pi^4 M_N))$ is Next-to-next-to-leading order (NNLO). Actually,
  as noted in
  Refs.~\cite{CalleCordon:2009ps,RuizArriola:2009bg,Cordon:2009pj} the
  expected large $N_c$ behaviour~\cite{Banerjee:2001js} does not
  hold for the (Two Pion Exchange) chiral potentials {\it even} after
  inclusion of $\Delta$~\cite{Kaiser:1998wa}. Likewise, the Wigner
  symmetry pattern is not fulfilled for those chiral
  potentials~\cite{CalleCordon:2008cz,CalleCordon:2009ps}.}.

\section{MESON EXCHANGE CURRENTS}

\begin{figure}[tbc]
\includegraphics[height=5.5cm,width=5.5cm,angle=270]{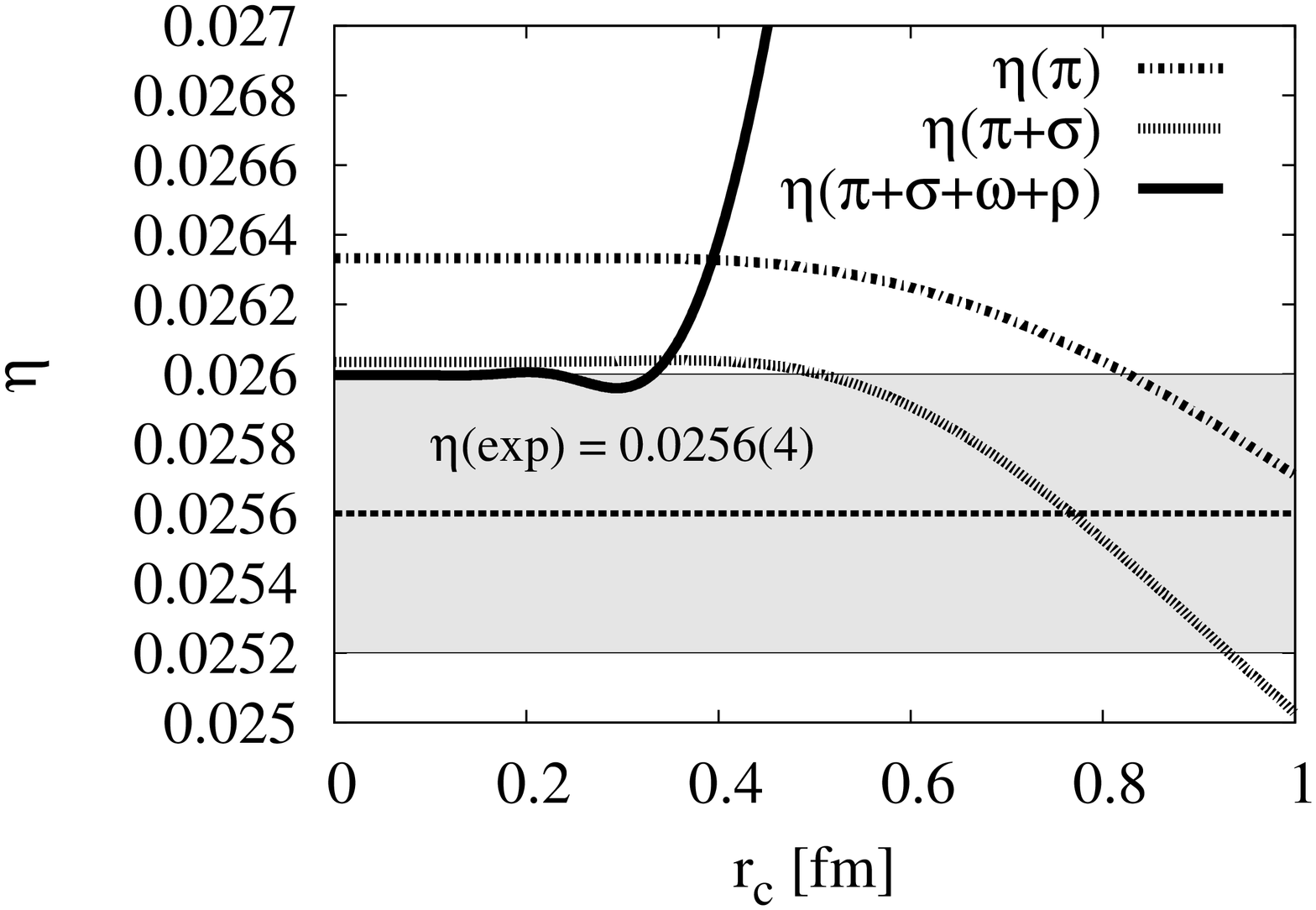}
\includegraphics[height=5.5cm,width=5.5cm,angle=270]{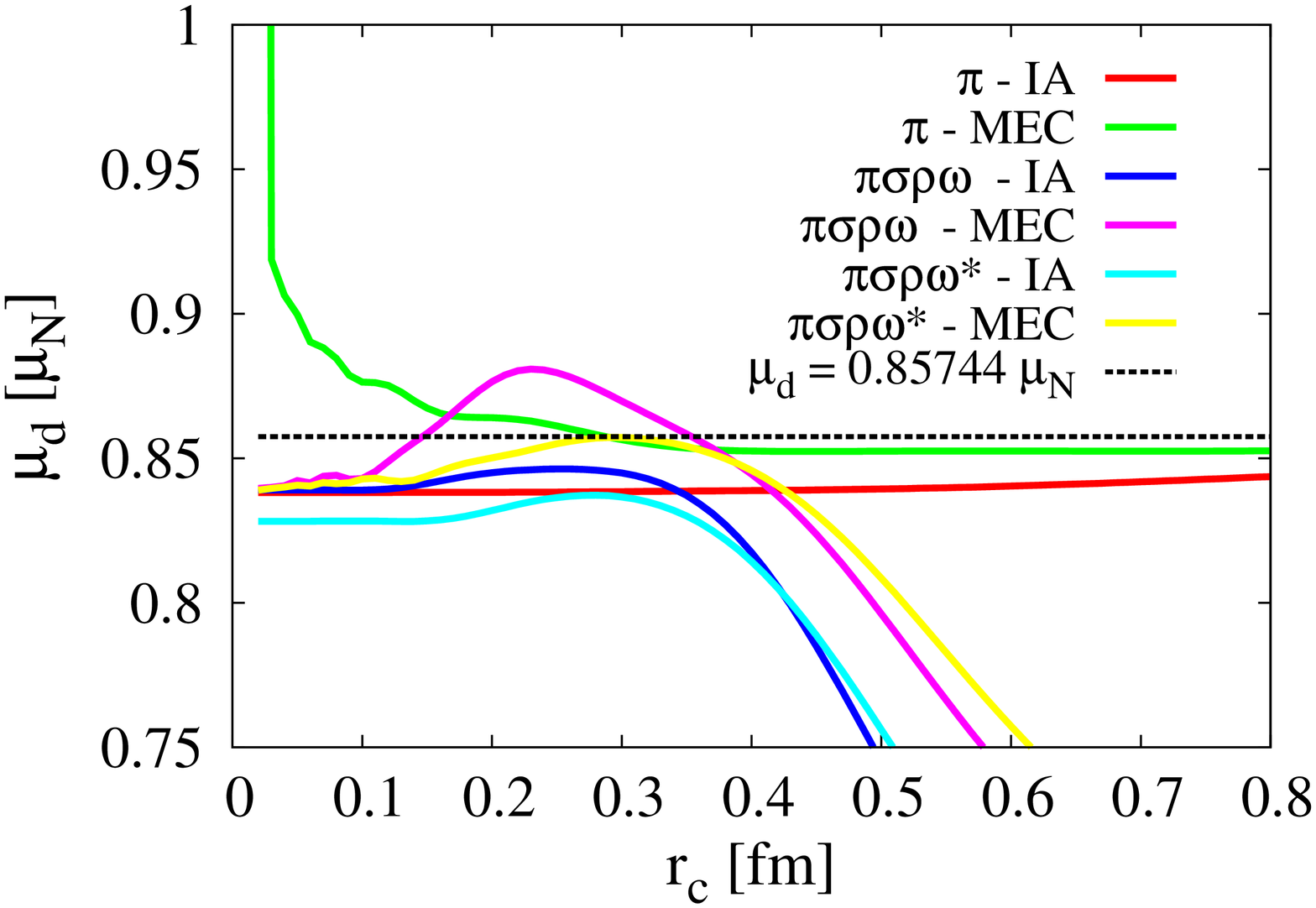}
\includegraphics[height=5.5cm,width=5.5cm,angle=270]{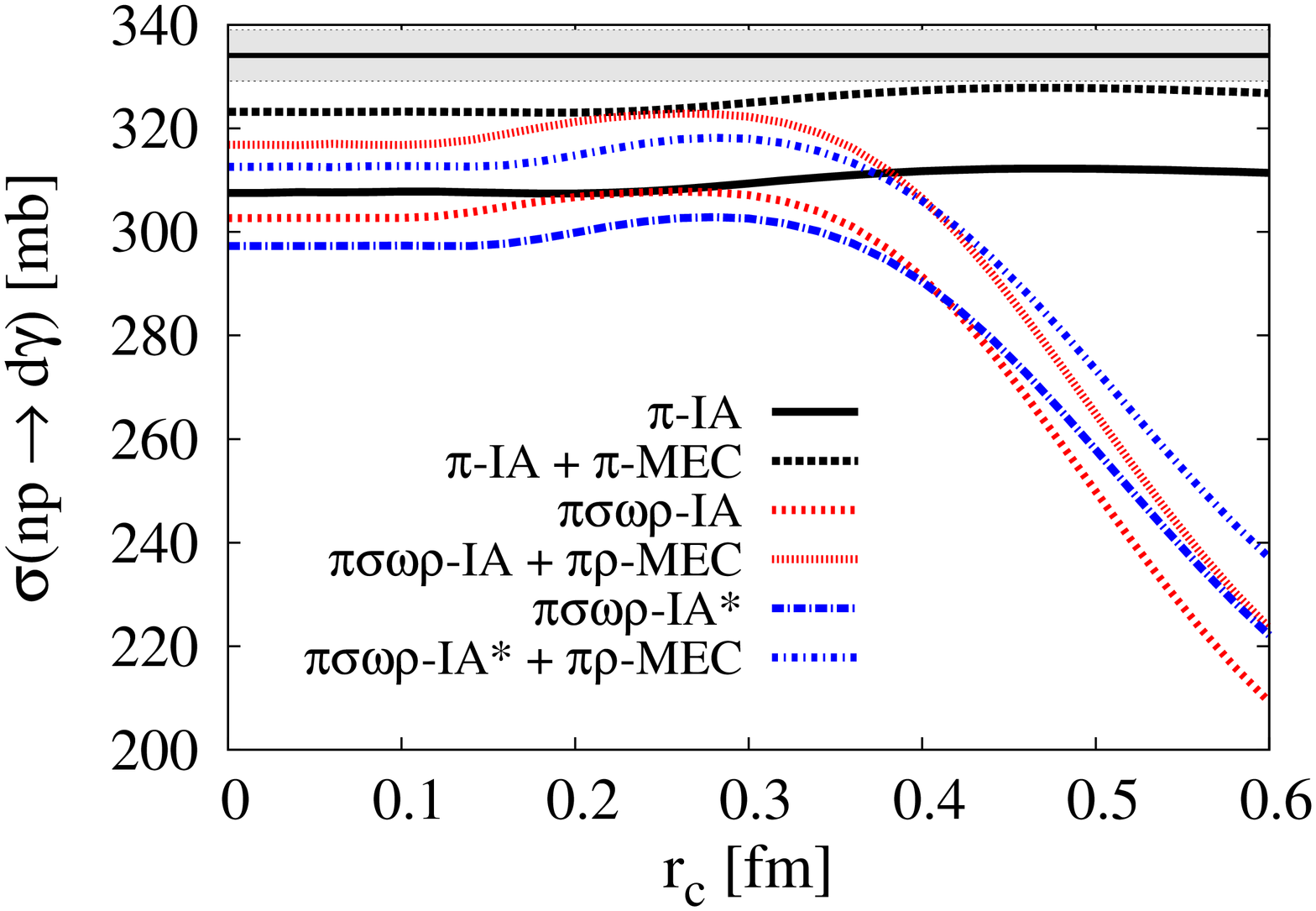}
\caption{Short distance Cut-off, $r_c$ (in fm), dependence of several
  observables including when necessary MEC in addition to Impulse
  Approximation (IA) compared with experimental bands.  Asymptotic D/S
  ratio $\eta =0.0256(4)$ (left). Deuteron magnetic moment $\mu_d=
  0.85744 \mu_N$ (middle).  Neutron capture cross section and
  $\sigma(np \to d\gamma) = 334.2(5) {\rm mb}$. (right)
}
\label{fig:mecs}
\end{figure}

Gauge invariance is easily preserved within our coordinate space
approach by keeping the {\it same} boundary condition as for the
potential in the large $N_c$ setup without need of new parameters (see
also Ref.~\cite{Riska:2002vn}). The simplest (purely transverse) MEC
correction to the deuteron magnetic moment in the Impulse
Approximation (IA) $ \mu_d^{\rm IA} = (\mu_p+\mu_n) + \frac32 \left(
\mu_p+\mu_n + \frac12 \right) P_D $ is shown in Fig.~\ref{fig:mecs}
(middle panel) as a function of $r_c$. Likewise, we show (right panel)
the neutron capture cross section. The (longitudinal) MEC
contribution yields a constant shift at relatively large distances.
The different short distance behaviour between transverse
and longitudinal MEC's will be elaborated elsewhere.

\section{Conclusions}

Self-adjointness of the two-body Hamiltonian and current conservation
are simply intertwined within the renormalization with boundary
conditions approach above a certain cut-off distance.  Current
conservation is guaranteed at any value of the cut-off as long as
matrix elements are consistently evaluated with NN wave functions
constructed from the meson exchange Hamiltonian. The conditions for
finiteness for both deuteron, scattering and electromagnetic matrix
elements of longitudinal MEC's coincide.

\begin{theacknowledgments}
This work is supported by the Spanish DGI and FEDER funds with grant
FIS2008-01143/FIS, Junta de Andaluc{\'\i}a grant FQM225-05, and EU
Integrated Infrastructure Initiative Hadron Physics Project contract
RII3-CT-2004-506078.
\end{theacknowledgments}

\bibliographystyle{aipproc}   


\end{document}